\documentclass[reprint,amsmath,amssymb,prl,floatfix]{revtex4-1}
\usepackage{graphicx}

\begin{document}

\title{Intrinsic unpredictability of strong El Ni\~no events}
\author{John Guckenheimer}
 \affiliation{Mathematics Department, Cornell University, Ithaca, NY 14853}
 \email{jmg16@cornell.edu}
\author{Axel Timmermann}
\affiliation{International Pacific Research Center, School  of Ocean and Earth Science and Technology, University of Hawaii at Manoa, Honolulu, Hawaii, 96822} 
\author{Henk Dijkstra}
\affiliation{Institute for Marine and Atmospheric research Utrecht, 
Department of Physics and Astronomy, Utrecht University, Utrecht, 
The Netherlands} 
\author{Andrew Roberts}
\affiliation{Mathematics Department, Cornell University, Ithaca, NY 14853}

\date{\today}

\begin{abstract}
 
The El Ni\~no-Southern Oscillation (ENSO) is a mode of interannual variability in 
the coupled  equatorial ocean/atmosphere Pacific. El Ni\~no describes 
a state in which sea surface temperatures in the eastern Pacific increase 
and upwelling of colder, deep waters diminishes. El Ni\~no events typically 
peak in boreal winter, but their strength varies irregularly on decadal time 
scales. There were exceptionally strong El Ni\~no events in 1982-83, 
1997-98 and 2015-16 that affected weather on a global scale. Widely publicized 
forecasts in 2014 predicted that the 2015-16 event would occur a year earlier.
Predicting the strength of El Ni\~no is a matter of practical concern 
due to its effects on hydroclimate and agriculture around the world.
This paper presents a new robust mechanism limiting the predictability of strong 
ENSO events: the existence of an irregular switching  between an oscillatory state  
that has strong  El Ni\~no events  and a chaotic state that lacks strong events, which 
can be induced by very weak seasonal forcing  or noise. 
\end{abstract}

\maketitle

\section{}

The problem in predicting  El Ni\~no events is that they occur quite  irregularly, and their 
development seems to be different each time \cite{McPhaden2015}. The  ENSO 
phenomenon is thought to be an internal mode of the coupled  equatorial ocean-atmosphere system  
which can be  self-sustained or excited by random  noise \cite{Federov2003}. The 
interactions of the internal mode  and the external seasonal forcing can lead to 
chaotic behavior through nonlinear 
resonances  \cite{Tziperman1994,Jin1994}. On the other hand, the dynamical 
behavior can be strongly influenced  by noise, in particular  westerly wind bursts   \cite{Lian2014}
which can either be viewed as additive \cite{Roulston2000a}  or 
multiplicative noise  \cite{Eisenman2005}. 
During boreal spring the 
coupled ocean-atmosphere system is thought to be at its frailest state 
\cite{Webster1992QJRMS}.  Then the system is most susceptible to 
perturbations \cite{Webster1995MAP} which leads  to a `spring' predictability 
barrier in April/May \cite{Latif1994CD}.  The role of the  initial error pattern has been 
emphasized and in particular its interaction with  the seasonal cycle and the internal 
cycle  \cite{Mu2007,Duan2009,Yu2012}. 

Processes determining ENSO variability  and its limits of predictability have been 
discovered with conceptual models \cite{Suarez1988,Tziperman1998,Jin1997}. Such 
models seek to identify key dynamical processes in the complex, coupled 
atmosphere/ocean system in the tropical Pacific. This paper investigates the predictability 
of El Ni\~no using the low-dimensional conceptual Jin-Timmermann (JT) model of ENSO 
originally proposed 
by Jin \cite{Jin1997} and then extended  by  Timmermann et al. \cite{tim03}. The state 
space variables of this deterministic  model are sea surface temperatures of the 
equatorial western Pacific and  eastern Pacific, and the thermocline depth (e.g. the 
depth of the 20$^\circ$C isotherm) of the 
Western Pacific. Nonlinear  terms in the model  represent feedbacks due to winds, 
ocean currents and thermocline dynamics that  affect these state variables. 
Comparisons of this reduced order model with  simulations of coupled global climate 
models confirm that the model embodies  the basic features of ENSO in the tropical 
Pacific \cite{tim01}. 

Recently,  a  multiple time scale analysis of the JT model  was performed to 
gain new  insight  into the dynamics  \cite{roberts2016}.  This work transformed the model  
equations into an equivalent dimensionless system whose equations are
\begin{equation}  \label{fast}
\begin{cases}
x' &=\rho \delta (x^2 - ax) + x \left( x+y+c - c \tanh (x+z) \right)\\
y' &= -\rho \delta (ay + x^2) \\
z' &= \delta (k-z- \frac{x}{2}), 
\end{cases}
\end{equation}
The variables $x,y$ and $ z$ of this  model represent the sea surface 
temperature difference between the eastern and western equatorial Pacific, sea 
surface temperature of the western equatorial Pacific relative to a nominal 
reference temperature, and 
thermocline depth of the western Pacific, respectively. Each of the five 
parameters $\delta,\rho,c,k$ and $a$ represents a combination of 
physical characteristics of the tropical Pacific. Time dependent effects from 
the climatological annual cycle of the air-sea heat flux 
\cite{SteinTimmermann2014} will be included to the model as
explored below.

\begin{figure}
\includegraphics[width=0.45\textwidth]{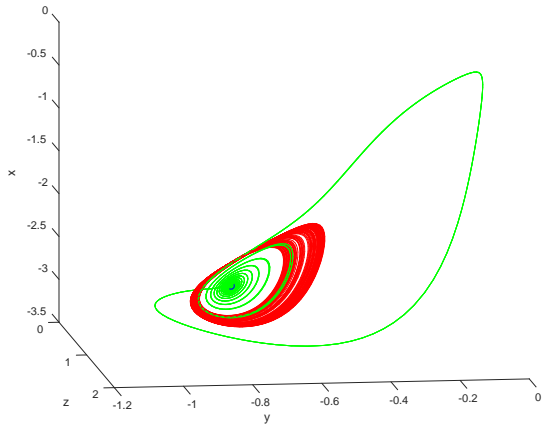}
  \caption{ Phase portrait of a chaotic attractor (red) and an MMO periodic orbit (green) that coexist at 
parameters  $[\delta,\rho,c,k,a] = [0.225423,0.3224,2.3952,0.4032,7.3939]$. 
The equilibrium point is a partially obscured blue dot.}
\label{enso_e_bistable_pp}
\end{figure}

This paper focuses attention on the behavior of the model with
parameters 
$$[\delta,\rho,c,k,a] = [0.225423,0.3224,2.3952,0.4032,7.3939]$$ where there 
is bistability between a periodic orbit and a chaotic attractor.  The green 
trajectory in Figure~\eqref{enso_e_bistable_pp} is a mixed mode oscillation (MMO) with period 
approximately 12.1 years and a single intense El 
Ni\~no event each period, which emerges after a series of growing interannual oscillations. 
The red trajectory is chaotic but the amplitude of 
its oscillations varies in a much narrower range than those of the MMO.

\begin{figure}
  \includegraphics[width=0.45\textwidth]{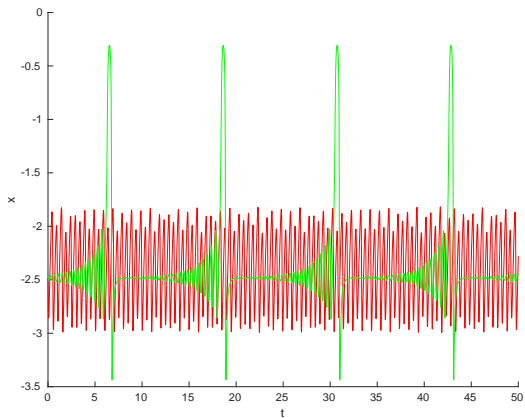}
  \caption{ Time series of the $x$ coordinate of the two trajectories shown in Figure~\eqref{enso_e_bistable_pp}. Units of time are years.
}
\label{enso_e_bistable_ts}
\end{figure}

A central feature of the 
MMO cycle is an equilibrium point that is a 
saddle focus: the MMO approaches the 
equilibrium along its one dimensional stable manifold and then spirals away,
following the two dimensional unstable manifold of the equilibrium. The number 
of oscillations 
of the MMO cycle and their minimum amplitude depends upon how close it 
approaches the equilibrium. The growing amplitude oscillations of the MMO cycle 
terminate in an intense El Ni\~no where the sea surface temperatures of eastern 
and western tropical Pacific are almost the same. This El Ni\~no is followed by 
a La Ni\~na event in which the system ``recharges'' \cite{Jin1997},
reestablishing 
higher sea surface temperatures and thermocline depth in the western than eastern Pacific. The 
recharged system flows back toward the equilibrium where the cycle repeats. 

The chaotic attractor is similar to those found in many other three dimensional 
vector fields. It can be analyzed by introducing a cross-section and studying 
its discrete time return map. The return map contracts a strip in one 
direction, and stretches and folds in the second direction. There is 
sensitivity to initial conditions within the chaotic attractor, but it is 
relatively mild: nearby initial conditions separate along the attractor without 
leaving it.

\begin{figure}
\includegraphics[width=0.45\textwidth]{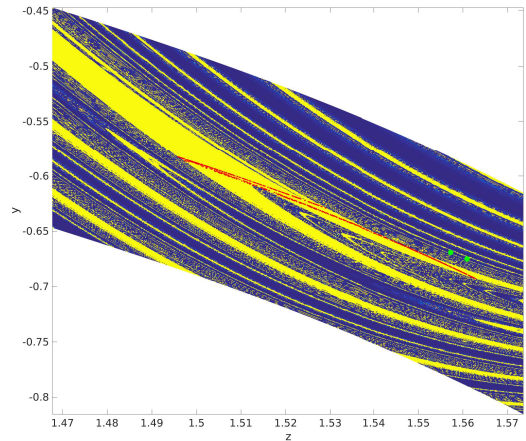}
  \caption{ The colored region is a 
500x500 grid that straddles the 
intersection of the chaotic attractor with the cross-section $x=-2.874$. Points 
are colored by whether the trajectories originating from the points approach 
the chaotic attractor (yellow) or the MMO (blue). The red markers and green 
dots 
mark intersections of the chaotic attractor and MMO with the cross-section, 
respectively 
}
\label{enso_e_imap}
\end{figure}

\begin{figure}
  \includegraphics[width=0.45\textwidth]{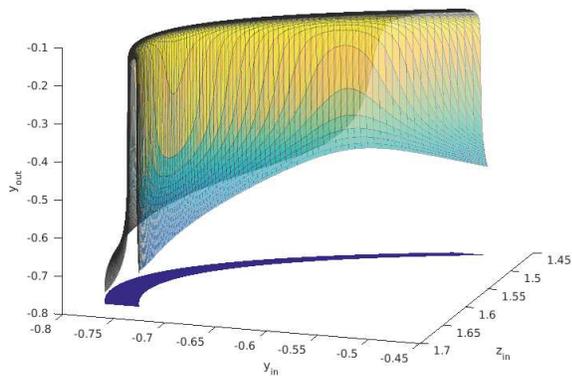}
  \caption{The image of the $y$-coordinate under the third iterate of the 
return map  of a 
thin strip of initial conditions (dark blue) of width 0.035 in the cross 
section. Points in the middle of the strip make a large amplitude 
excursion representing a strong El Ni\~no. The map stretches the $y$-coordinate 
by a factor of order $10^3$ where the returns have intermediate amplitude.}
\label{enso_e_sdic} 
  \end{figure}

The basins of attraction of the two attractors are intertwined with each other 
in an intricate way. They display fractal basin boundaries 
\cite{grebogi87}.
Even though 
almost all initial conditions presumably approach one of the two attractors, 
there are large regions in which it is difficult to predict which basin a 
chosen initial condition will belong to (Figure \eqref{enso_e_imap}).
Note that both attractors are very close to their basin boundaries. This 
suggests that there are nearby trajectories that separate abruptly. This is 
displayed in Figure \eqref{enso_e_sdic} which plots the third return of the 
$y$ coordinate in a strip of width 0.035 which spans points that lie in the 
basin of attraction of the MMO 
The plateau of this map consists of points that make a large amplitude oscillation
and tend to the MMO attractor. On either side of the basin, 
the relative distances between points of the steep region are expanded by the return 
map by as much as $10^3$: there is very high sensitivity to initial conditions 
here. This prompts the question as to whether small modifications of the system 
cause switching between the two attractors by creating trajectories that visit
the plateau of this (third) return map on some returns but not other. 

\begin{figure}
  \includegraphics[width=0.45\textwidth]{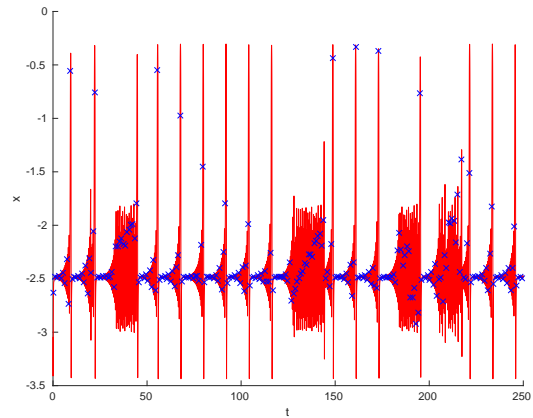}
  \caption{Time series for the x coordinate of the model with the parameter 
$a$ modified to be the periodic 
function of time $a = 7.3939+0.001\sin(1.8 t$).  The frequency and 
the units of time are years. The crosses are points at a specific phase of 
the forcing.
}
\label{enso_f_ts}
\end{figure}

\begin{figure}
  \includegraphics[width=0.45\textwidth]{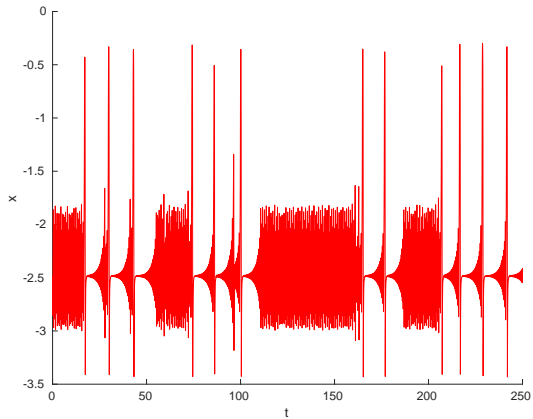}
  \caption{Time series for the x coordinate of the model with the parameter 
$a$ modified to be stochastic. With discrete time steps of approximately one day,
$a = 7.3939+0.001*N$ where $N$ is a sample path of the normal distribution.  
}
\label{enso_stoch_a}
\end{figure}

To investigate switching between the attractors, the model 
was modified by making the parameter $a$ a sinusoidal function of time with a 
period of one year to mimic  effects of the annual solar cycle \cite{tim03,SteinTimmermann2014}. 
Figure \eqref{enso_f_ts} shows the results of setting the amplitude of the periodic forcing 
to 0.001. Even at this small forcing amplitude, the system switches erratically 
between its two modes of variability. In a longer simulation of 6000 years, the 
largest number of consecutive MMO cycles was 24 and the smallest number was 1. 
Some epochs of chaotic oscillations were over 200 years in duration. There is 
no apparent pattern to these durations, so it seems that the timing of strong El 
Ni\~no events is unpredictable on a decadal time cycle. No attempt was made to 
identify precursors to these events, but that is clearly a matter of practical 
interest. Similar results can be obtained by including weak multiplicative 
forcing of the parameter $a$ instead of periodic forcing as shown in 
Figure \eqref{enso_stoch_a}.


%

Reconstructions of past climate from historical data indicate that ENSO is 
highly variable with long periods of larger and smaller variance on decadal 
time scales~\cite{cobb03,McGregorTimmermannClimateofPast2013}.
The above results demonstrate the existence of a new mechanism limiting 
predictability  of strong ENSO events. This  extreme variability appears 
in a simple low-dimensional  model that incorporates  coupling of basin-wide 
atmospheric and oceanic  waves and adjustment. Rapid divergence of nearby  
trajectories from one another  is concentrated in small regions of the state  
space of the model. If more detailed ENSO models behave similarly, this 
finding of bistability  implies strong  limits to the longer-term predictability
of strong El Ni\~no events. This needs to be considered when developing  
strategies for improving ENSO forecasts.

\section*{Methods}

Numerical simulations of the models in this paper were performed using an 
implementation of the algorithm DOP853~\cite{HNW} in MATLAB. Stringent error 
tolerances (typically $10^{-12}$) were used for calculating trajectories and 
events (intersections of a trajectory with a specified cross-section).

The parameters used in this study were located in the following manner. 
The program MATCONT \cite{matcont} was used to locate points of Hopf 
bifurcation and then to identify points of ``generalized Hopf'' bifurcations 
where  the bifurcation is neither subcritical nor supercritical. Starting at 
parameters with a supercritical Hopf bifurcation and varying $\delta$, periodic 
orbits were continued until they bifurcated through a period doubling cascade 
into a chaotic attractor. Further increases in $\delta$ led to a regime with an 
MMO attractor. Decreasing $\delta$ showed that the ranges of $\delta$ with MMO 
and chaotic attractors overlapped. The value of $\delta$ used in this study is 
in the overlap region.

The red and green markers in Figure \eqref{enso_e_imap} were obtained by 
computing intersections of trajectories in the two attractors with the 
cross-section $x=-2.874$. (This is the value of $x$ at the equilibrium of the 
system.) A quadratic function $y = q(z)$ was fit with least squares (MATLAB 
command polyfit) to the intersection points of the chaotic attractor and the 
cross-section. The graph of this quadratic function was translated in the $y$ 
direction to produce a 500x500 grid of initial conditions used in producing the 
figure. At each point in the grid, a trajectory was computed and then the point 
was colored with a value corresponding to the maximum value of $y$ in its 
intersections with the cross-section. When the trajectory approaches the MMO, 
this maximum comes from the strong El Ni\~no events and is larger than the 
values obtained from the chaotic attractor.

The strip of initial conditions used in computing Figure \eqref{enso_e_sdic} 
was found by integrating initial conditions in the cross-section with $y=-0.4$ 
backward to their third return to the cross-section, fitting these points by a 
cubic curve and then translating this curve in the $y$ direction. The third 
return to the cross-section was computed on a 100x100 grid in this region. The 
figure plots the value of $y$ at the endpoint as a function of the initial 
point. The use of the third return in these calculations was motivated by the 
computation of an unstable periodic orbit in this region which intersects the 
cross-section three times.

\begin{acknowledgments}
{\bf Acknowledgments:}  The research of John Guckenheimer was partially supported by NSF 
Grant 1006272.  Henk Dijkstra's work was partially supported by a Mary Shepard B. Upson 
Visiting Professor position at the College of Engineering of  Cornell University, Ithaca, 
NY. He thanks for Prof. Paul Steen (Cornell University) for being his host and the many 
interesting discussions. Axel Timmermann was supported by NOAA grant NA15OAR4310117. 
\end{acknowledgments}

\bibliographystyle{apsrev}
\bibliography{enso_bistability_prl}

\end{document}